\documentclass{article}
\usepackage{arxiv}

\usepackage[utf8]{inputenc} 
\usepackage[T1]{fontenc}    
\usepackage{hyperref}       
\usepackage{url}            
\usepackage{booktabs}       
\usepackage{amsfonts}       
\usepackage{nicefrac}       
\usepackage{microtype}      
\usepackage{natbib}
\usepackage{amsmath,graphicx,xcolor}
\usepackage{caption,subcaption}
\usepackage[normalem]{ulem}
\usepackage{amsthm}
\usepackage{hyperref}
\usepackage{amsmath,graphicx,xcolor}
\usepackage[normalem]{ulem}
\usepackage{bm}

\title{How to ``Improve" Prediction \\ Using Behavior Modification}

\author{
Galit Shmueli \\
  Institute of Service Science\\
  National Tsing Hua University\\
  Hsinchu 30013, Taiwan \\
  \texttt{galit.shmueli@iss.nthu.edu.tw} \\
\And 
Ali Tafti \\
  College of Business Administration\\
  University of Illinois at Chicago\\
  Chicago, Illinois 60607 \\
  \texttt{atafti@uic.edu} \\
}

\begin{document}
\maketitle

\begin{abstract}
Many internet platforms that collect behavioral big data {use it to} predict user behavior for internal purposes and for their business customers 
(e.g., advertisers, insurers, security forces, {governments,} political consulting firms) who utilize the predictions for personalization, targeting, and other decision-making. {Improving predictive accuracy is therefore extremely valuable.} Data science researchers design algorithms, models, and approaches to improve prediction. Prediction is also improved with larger and richer data. Beyond improving algorithms and data, platforms can stealthily achieve better prediction accuracy by pushing users' behaviors towards their predicted values, using behavior modification techniques, thereby demonstrating more certain predictions.  
Such apparent “improved” prediction can 
result from employing reinforcement learning algorithms that combine prediction and behavior modification. This strategy is absent from the machine learning and statistics literature. Investigating its properties requires integrating causal with predictive notation. To this end, we incorporate Pearl's causal $do(.)$ operator into the predictive vocabulary. We then decompose the expected prediction error given behavior modification, and identify the components impacting predictive power. Our derivation elucidates implications of such behavior modification to data scientists, platforms, their customers, and the humans whose behavior is manipulated. Behavior modification can make users' behavior more predictable and even more homogeneous; yet this apparent predictability might not generalize when business customers use predictions in practice. Outcomes pushed towards their predictions can be at odds with customers' intentions, and harmful to manipulated users.
\end{abstract}


\section{Introduction: {Prediction, Prediction Products, and Behavior Modification}}
\label{sec-intro}

Recent years have seen an incredible growth in predictive modeling of user behavior using behavioral big data in both industry and in academia. Behavioral big data are large and highly detailed datasets on human and social actions and interactions \citep{shmueli2017research}. 
Predictions {based on such data} now shape almost every aspect of modern life \citep{agrawal2018prediction}. Internet platforms, such as Google and Facebook, predict user behavior for internal purposes and for their customers who utilize the predictions for personalization, targeting, and other decision-making. Predicted behaviors of interest to platforms and their customers include the probability of purchase, churn, engagement, and even {behaviors such as} voting intentions and life events such as pregnancy\footnote{\url{www.nytimes.com/2012/02/19/magazine/shopping-habits.html}}. Henceforth in this paper, we use the term \emph{customers} to refer to the platform's business customers, which we distinguish from the platform's \emph{users}, whose behavior is subject to prediction and modification.

\subsection{Prediction Products}
In \emph{The Age of Surveillance Capitalism}, \citet{zuboff2019age} describes the processes used by several large internet platforms that collect behavioral big data to package the raw material of users' actively shared data and passively generated data (e.g.~location data, 
friendship ties, device information) into \emph{prediction products}, which are then sold to business customers such as insurance companies, marketers, advertisers, security forces, governments, and political consulting firms. {Prediction products ``anticipate what you will do now, soon, and later" \citep[][p. 8]{zuboff2019age}.}
{Such} predictions are typically used by platform customers to modify and shape users' behavior toward desired commercial or other outcomes.\footnote{\url{www.theguardian.com/technology/2019/jan/20/shoshana-zuboff-age-of-surveillance-capitalism-google-facebook}}
\footnote{{\citet[][p.15]{zuboff2019age} {explains how} these prediction products are traded in a new kind of marketplace for behavioral predictions that she calls \emph{behavioral futures markets}. As \citet[][p. 566]{andrew2021general} {elaborate}, the valuable combination of ``data, predictive algorithms, and behavioral modification techniques" creates {such} an emerging behavioral futures market. For example, Google’s “clickthrough rate” was the first globally successful prediction product, and its ad markets were the first to trade in human futures (\url{https://www.project-syndicate.org/onpoint/surveillance-capitalism-exploiting-behavioral-data-by-shoshana-zuboff-2020-01})}.} 

One example of a prediction product is the recently launched Google Analytics \emph{predictive metrics} service that ``automatically enriches your data by bringing Google machine-learning expertise to bear on your dataset to predict the \emph{future behavior} of your users".\footnote{``Purchase Probability, which predicts the likelihood that users who have visited your app or site will purchase in the next seven days\ldots Churn Probability, predicts how likely it is that recently active users will not visit your app or site in the next seven days."  \url{https://blog.google/products/marketingplatform/analytics/new-predictive-capabilities-google-analytics/} archived at \url{https://archive.ph/Zms2O}}
Another example is Facebook's \emph{loyalty prediction} service that offers advertisers the ability to target users based on how they \emph{will} behave, what they \emph{will} buy, and what they \emph{will} think\footnote{{\url{https://theintercept.com/2018/04/13/facebook-advertising-data-artificial-intelligence-ai/}}}.

\citeauthor{zuboff2019age}'s (\citeyear{zuboff2019age}) book, which sounds an alarm about the increasingly intrusive and exploitative practices of digital platform firms, has been received with both acclaim and criticism. Her work takes a condemnatory stance towards the major digital platforms, arguing that their practices are trending not just towards influence but pervasive control of human behavior, and thus pose a great danger to the autonomous and lived experience of humanity. We take 
an analytical approach, by considering the potential scope and effects of behavioral modification should digital platforms choose to engage in such strategies. Regardless of one’s disposition towards \citet{zuboff2019age}'s arguments, there is a need for better understanding and careful analysis of how the scenarios depicted in that work might play out in practice. Hence, we focus on prediction products and the strategies that platforms might take to capture their place in this market. Essentially, the more accurate these prediction products, the higher value they provide {to the platforms'} customers, and in turn the higher the revenues for the platforms. Platforms might also compete for customers of their prediction products. Hence, platforms have a strong incentive to improve prediction accuracy.

\subsection{Modifying User Behavior}
Digital platforms now routinely use \emph{behavior modification} (BMOD) techniques to change their users' behaviors both online and offline. Behaviors include clicking an ad, purchasing an item, posting sensitive information, visiting a doctor, voting, and more.
BMOD techniques are classically defined as ``an observable, replicable and irreducible component of an intervention designed to alter or redirect causal processes that regulate behavior" \citep{michie2013behavior}. BMOD techniques derive from principles of behaviorist psychology and include nudging, herding, and operant conditioning, among others.
The most popular technique is the \emph{nudge}, defined as ``any aspect of the choice architecture that alters people's behavior in a predictable way without forbidding any options or significantly changing their economic incentives" \citep[][p. 6]{thaler2009nudge}. Designers of choice environments take advantage of human cognitive limitations to manipulate the choice environment to subtly guide behavior by gently nudging them toward certain choices \citep{schneider2018digital}. \citet{zuboff2019age} identifies two more types of behavior modification: \emph{herding}, which is controlling key elements in a person's immediate context in order to guide their behavior towards a predictable one; and \emph{operant conditioning}, a term coined by the famous behavioral psychologist BF Skinner, which uses positive and negative reinforcement to encourage certain behaviors and extinguish others. 
When implemented on platforms, BMOD techniques are known as \emph{persuasive technology} \citep{fogg2002persuasive}, and can be used to adaptively and automatically tailor behavioural interventions to exploit unique psychological characteristics and motivations. Platform BMOD is implemented via various machine learning algorithms that operate in a data-driven, autonomous, interactive, and sequentially-adaptive manner \citep{greene2022barriers}. 
Behavioral interventions range in their transparency: some are visible to users, such as chatbots, suggestions by recommender systems, and app notifications, while others are less so, such as  A/B testing\footnote{{Randomized experiments, including the kinds of A/B tests done by digital platforms, impose interventions on a random subset of users in order to evaluate the effect of those interventions.}}, feed filtering, comment moderation on social networks, and deceptive interface design choices \citep{mathur2019dark}. Platforms employ BMOD to provide personalized services, increase user engagement, ``hook" users by habit formation \citep{eyal2014hooked}, and generate further {behavioral big data}, among other purposes. Stanford university's Behavior Design Lab director BJ Fogg lists seven types of \emph{persuasive technology} tools \citep{fogg2002persuasive}. While the field of marketing has used behavior modification even prior to the advent of the internet \citep{nord1980behavior}, today's technologies and big data enable more covert, pervasive, and powerful manipulation due to their networked, continuously updated, dynamic and pervasive nature \citep{yeung2017hypernudge}. 
\citet{zuboff2019age} explains,
\begin{quote}
    ``These interventions are designed to enhance certainty by doing things: they nudge, tune, herd, manipulate, and modify behavior in specific directions by executing actions as subtle as inserting a specific phrase into your Facebook news feed, timing the appearance of a BUY button on your phone, or shutting down your car engine when an insurance payment is late." (p. 200)
\end{quote}
While these examples do not necessarily involve prediction, prediction-based behavior modification is common in recommendation systems, targeted advertising, precision marketing, and other personalized interventions intended to cause human users to change their behavior in a specific direction that is beneficial to the intervention initiator: towards longer online engagement, higher purchase propensity, or increased information sharing.

\subsection{Combining Prediction and Behavior Modification}
{Our focus is on the new capability that results from combining prediction and BMOD strategies that are now available to platforms. \citet{zuboff2019age} explains how a platform that uses prediction and behavior modification for its own commercial gain can be in conflict with users' well-being and agency. Engineering decisions made to maximize profitability can lead to changes in social systems that are drastic, opaque, effectively unregulated, and massive in scale \citep{bak2021stewardship}. 
Historian Yuval Noah Harari suggests that such a combination of powers provides the capability to ``hack" human beings\footnote{The TED Interview: ``Yuval Noah Harari reveals the real dangers ahead" \url{https://www.ted.com/talks/the_ted_interview_yuval_noah_harari_reveals_the_real_dangers_ahead}, minute 34:15}:
\begin{quote}
    “The ability to hack human-beings means the ability to understand humans better than they understand themselves. Which means being able to predict their choices\ldots to manipulate their emotions, to make decisions for them.”
\end{quote}
}

{Scholars from computer science, behavioral science, media studies, and law have pointed out the strong incentive for platforms to ``make predictions true".  We are interested in studying the possibilities of ``improving" prediction by using BMOD, in the sense of ``making predictions come true".}

Commenting on Facebook's loyalty prediction  service,\footnote{\url{https://theintercept.com/2018/04/13/facebook-advertising-data-artificial-intelligence-ai/}} 
Law Professor Frank Pasquale said he
worried how the company could turn algorithmic predictions into ``self-fulfilling prophecies,” since ``once they've made this prediction, they have a financial interest in making it true.”\footnote{Further quoting Frank Pasquale: ``That is, once Facebook tells an advertising partner you’re going to do some thing or other next month, the onus is on Facebook to either make that event come to pass, or show that they were able to help effectively prevent it (how Facebook can verify to a marketer that it was indeed able to change the future is unclear)."}
 Computer scientist and AI expert Stuart Russell describes how platforms improve prediction by changing users' preferences so they become more predictable \citep{russell2019human}.\footnote{ 
    ``Content selection algorithms on social media\ldots are designed to maximize click-through. The solution is simply to present items that the user likes to click on, right? Wrong. The solution is to change the user’s preferences so that they become more predictable." \citep{russell2019human}}
Media theorist Douglas Rushkoff explains that the better the platform is able to make users conform to their algorithmically determined destiny, the more it can ``boast both its predictive accuracy and its ability to induce behavior change" \citep[][p. 69]{rushkoff2019team}.
 
Examining the combined prediction and BMOD capabilities from a data science point of view, we study how such a combined strategy optimized to minimize prediction error could affect the different stakeholders: the platform, its customers purchasing prediction products, and the platform users. By describing the scenario in statistical terms, we show that aiming to minimize prediction error can result in misleading platform customers and manipulating humans in possibly dangerous directions. This is especially the case when predicting {\it unwanted or risky behaviors}, in the form of risk scores. An extreme example is predicting mental health risk for a healthcare stress reduction app. While the app maker aims to lower stress of high risk users, the platform can achieve low prediction error by turning high risk predictions into high risk realities.

The goal of this work is not to offer new evidence nor to characterize all the ways in which digital platforms engage in behavioral modification strategies. Given that some firms have already demonstrated both the capabilities and incentives to implement various forms of behavior modification, our goal rather is to introduce a technical vocabulary and notation that enables investigation of such strategies. Technical terminology and notation are needed in order to identify the properties and implications of the behavior modification approach to resulting predictive power. 
The thought process gives rise to various questions that we shall consider here and that future research can investigate in even more detail, including: Can behavior modification mask poor predictive {capabilities}? Can one infer from the manipulated predictive power  the counterfactual of non-manipulated predictive power? Can customers running routine A/B tests on the platform detect this scheme? What are the roles of personalized predictions and of personalized behavior modifications within the error minimization strategy?

Our goal is to make transparent the effects of behavior modification on predictive power, thereby enabling the study of its impact on business, social, and humanistic aspects, as well as potential implications. In order to enable the analysis and evaluation of the effect of behavior modification on predictive power, we use the $do(.)$ operator by \citet{pearl2009causality}.   
This operator is useful for our approach because it expresses manipulation of a variable within a system of causal relations. While $do$ calculus is well developed for causal effects identification \citep{pearl2009causality}, the challenge here lies in combining the causal $do(.)$ operator into the existing correlation-based predictive framework.

The paper proceeds as follows. In Section \ref{sec-pred} we describe the traditional statistical and machine learning approach to reducing prediction error. Section \ref{sec-new-pred} introduces the new approach to reducing prediction error that relies on behavior modification. In Section  \ref{sec-formalization} we formalize the approach using statistical language and notation and analyze its implications. Section \ref{sec-discuss} discusses implications, from technical and business implications to humanistic and societal. Section \ref{sec-conclude} provides conclusions.

\section{Reducing Prediction Error by Improving Prediction}
\label{sec-pred}
The fields of machine learning and statistics have been introducing new and improved models, algorithms, approaches, and even data, aimed at improving predictive power. Approaches such as regularization, boosting, and ensembles have proven highly useful in generating more precise predictions. From transparent regression models and tree-based algorithms, to more blackbox support vector machines, k-nearest neighbors, neural nets and especially deep learning algorithms, their justification and adoption lies in their ability to capture intricate signals linking inputs and a to-be-predicted output. 

Predictive performance is typically measured by out-of-sample prediction errors, which compare predicted values with actual values for new observations. More formally, the prediction error $e_i$ for record $i$ is defined as the difference between the actual outcome value $y_i$ and its prediction $\hat{y}_i$, that is $ e_i = y_i - \hat{y}_i.$
For a sample of $n$ records, we have a set of actual outcome values $\bm{y}=\left[y_1,y_2,\ldots, y_n\right]$, a set of predicted outcome values $\bm{\hat{y}}=\left[\hat{y}_1, \hat{y}_2, \ldots, \hat{y}_n\right]$, and a set of prediction errors $\bm{e}=\left[e_1,e_2,\ldots, e_n\right]$. For each record $i$ we also have predictor information in the form of $p$ measurements $\bm{x}_i = \left[x_{i,1},\ldots,x_{i,p}\right]$. The predictor information for $n$ records is contained in the matrix $\bm{X}$.
Predicted values are obtained from $\hat{f}$, the {models trained on a dataset of} inputs $\bm{X}$ and actual outcomes $\bm{y}$, so that $\bm{\hat{y}}=\hat{f}(\bm{X})$.\footnote{We assume a prediction is made at time $t$ for an outcome $y$ that will occur at a future time $t+k$ ($k>0$). For record $i$, this can be written as $\hat{y}_i{(t+k})|\bm{x}_i(t)$. However, for simplicity of exposition, we drop the time indexes.}

\subsection{Targets of statistical and machine learning efforts}
{Machine learning} algorithms and {predictive statistical models}\footnote{{The term \emph{model} has different meanings in statistics and in machine learning. We use the term \emph{model} (or \emph{predictive model}) in the statistical sense of representing the relationship between an outcome and predictors, denoted by $f$, independent of a particular dataset. We use \emph{estimated model} or \emph{trained model}, denoted $\hat{f}$, to refer to the model after being trained on data. We use the term \emph{algorithm} for the procedure applied to the data that yields $\hat{f}$. For example $y=\beta_0+\beta_1 X + \epsilon$ is a model, which can be trained on data using a maximum likelihood algorithm, resulting in the estimated model $\hat{y}=5+3X$. In machine learning, \emph{model} refers to the result of training an algorithm on data (equivalent to  \emph{estimated model} when a statistical model exists). For example, a ``classification tree model" in machine learning is the result of applying some tree algorithm to data. 
}} 
are designed and tuned to minimize some aggregation of the error values ($\bm{e}$) by operating on the predicted values ($\bm{\hat{y}}$). Improving predicted values is typically achieved by improving three components:
\begin{enumerate}
    \item The structure of $\hat{f}$ that relates the predictor information $\bm{X}$ to the outcome, 
    and methods), 
    \item the estimation/computation of $\hat{f}$ {(e.g., {through better} algorithms)}, and 
    \item the quality and quantity of {data} ($\bm{X}$ and $\bm{y}$).  
    Larger, richer behavioral datasets have been shown to improve predictive accuracy \citep{martens2016mining}.
\end{enumerate}
Towards this end, companies such as Google, Facebook, Uber, Netflix, and Amazon have been investing in improving predictions through collecting, buying, storing and processing unprecedented amounts and types of data. They have also hired top data science talent, acquired AI companies, and developed in-house predictive, algorithms, platforms, and computational infrastructure.

\begin{figure}
    \includegraphics[width=\textwidth]{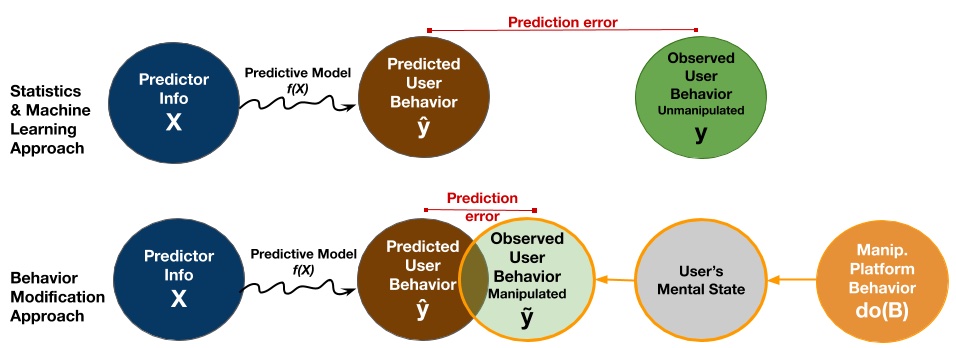}
    \caption{Prediction error with no behavior modification (top) vs.~with behavior modification (bottom). Manipulating platform behavior $do(B)$ pushes the observed user behavior towards its predicted value. Note that only orange arrows denote causal effects; squiggly black arrows denote a correlation-based predictive relationship.}
    \label{fig:error-diagram}
\end{figure}

 In all these approaches, the actual outcome values $\bm{y}$ are considered fixed {as if they represent unmanipulated outcomes}. The top panel of Figure \ref{fig:error-diagram} illustrates the statistical and machine learning approach of improving the above three components in order to minimize prediction error.

\subsection{Components affecting predictive power: dissecting the expected predicted error}
When predicting an outcome $y$ that is not expected to be manipulated between training and deployment, we anticipate prediction error due to the inability of the model $\hat{f}$ to (1) correctly capture the underlying $f$ even with unlimited training data (bias), (2) correctly estimate $f$ due to insufficient data (variance), (3) capture the errors for individual observations $\bm{\epsilon}$ (noise).
For predicting a numerical outcome or probability for a new observation, these three sources are formalized through a bias-variance decomposition of the expected prediction error (EPE), using squared-error loss\footnote{Assuming underlying model $E[y|\bm{x}]=f(\bm{x})+\epsilon$, where  $\epsilon$ has zero mean and variance $\sigma^2$.} 
\citep{geman1992neural}:
\begin{equation}
\begin{split}
EPE(\bm{x}) &= E\left[ \left((y|{\bm{x}}) - \hat{f}(\bm{x})\right)^2 \right]\\
&= E\left[\epsilon^2\right] + \left( f(\bm{x}) - E[\hat{f}(\bm{x}] \right)^2 + E\left[\left(\hat{f}(\bm{x}) - E[\hat{f}(\bm{x}]\right)^2\right]\\
& = \sigma^2 + \left[Bias(\hat{f}(\bm{x}))\right]^2 + Var(\hat{f}(\bm{x})).
\end{split}
\label{eq:epe-non-manip}
\end{equation}

In statistics and machine learning, prediction is based on an assumption of continuity, where the predicted observations come from the same underlying processes and environment as the data used for training the predictive {model} and testing its 
predictive performance.
The deterministic underlying function $f$ and the random noise distribution are both assumed to remain unchanged  between the time of model training and evaluation and the time of deployment. This assumption underlies the practice of randomly partitioning the data into separate training and test sets (or into multiple \emph{folds} in cross validation), where the model is trained on the training data and evaluated on the separate test data. Of course, the continuity assumption is often violated to some degree depending on the distance (temporal, geographical, etc.) between the training/test data and the to-be-predicted data and how fast or abruptly the environment changes between these two contexts. These \emph{dataset shift} challenges \citep[see e.g.][]{moreno2012unifying, quionero2009dataset} can increase prediction errors beyond the disparity observed between training and test prediction errors. Hence, predictive power based on the test data might provide an overly optimistic estimate of
actual performance at deployment. 

We note that EPE being an expected value indicates that it is a population quantity. In practice, it is typically estimated using the mean of the squared prediction errors (MSE) in the test dataset.

\section{Reducing Prediction Error by Modifying User Behavior (``Improving" Prediction)}
\label{sec-new-pred}

Platforms now have the incentive and technology\footnote{For example, Facebook’s \emph{AI backbone} FBLearner Flow combines machine learning and experimentation capabilities that can be applied to the entire Facebook userbase \url{https://engineering.fb.com/core-data/introducing-fblearner-flow-facebook-s-ai-backbone/} archived at \url{https://archive.ph/P4kkE}} to minimize prediction errors in a direction that is absent from academic prediction research: by modifying {\it actual} behavior ($\bm{y}$).

\begin{figure}
    \centering
    \includegraphics[width=0.8\textwidth]{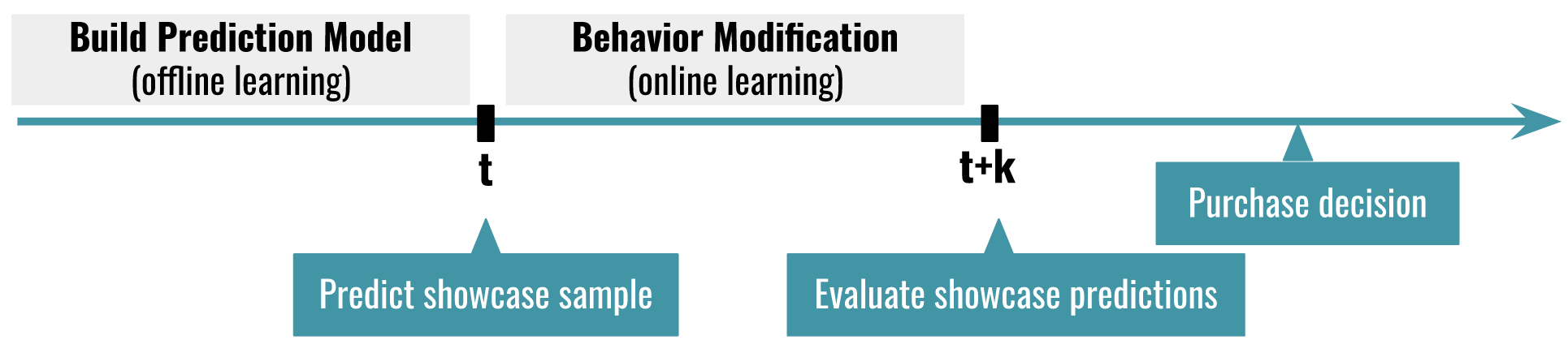}
    \caption{Sequence of events: prediction at time $t$ is followed by BMOD during $(t,t+k)$. After evaluating predictive accuracy at time $t+k$, the customer decides whether to purchase the platform's prediction product.}
    \label{fig:timeline}
\end{figure}

BMOD techniques can be used to minimize prediction error by pushing users' behaviors towards their predicted values, thereby ``improving" apparent prediction capability. This can be achieved via a predict-then-modify process: 
\begin{description}
    \item[Predict:] At time $t$, predict a user's future behavior at time $t+k$ ($k>0$).
    \item[Modify:] During period $(t,t+k)$, modify the user's behavior towards the predicted value in the course of continual monitoring of the user's behavior.
\end{description}
This sequence is illustrated in Figure \ref{fig:timeline} for a sample of users (\emph{showcase sample}). At time $t+k$ the outcome is observed, allowing the platform and customer to evaluate predictive accuracy and subsequently make purchase decisions.

The two enabling mechanisms for such a process are that (1) platforms have a plethora of powerful and tested BMOD tools, and (2) BMOD techniques are designed to modify behavior \emph{in a predictable way} -- here pushing outcome values towards their predicted values -- in order ``to shape individual, group, and population behavior in ways that continuously improve their approximation to guaranteed outcomes" \citep[][p. 339]{zuboff2019age}. We now describe a powerful machine learning method -- reinforcement learning -- that can be used to implement this strategy.

\subsection{Behavior modification via reinforcement learning}

The financial incentives and technical capabilities of internet platforms might entice platform data science teams under pressure to showcase predictive performance to engage in this prediction ``improvement"  strategy either knowingly or myopically. A technology that makes this especially suited towards this end is \emph{reinforcement learning} (RL) now often  used for implementing BMOD by platforms \citep{den2020reinforcement}.  

\begin{figure}[h]
    \centering
    \includegraphics[width=0.6\textwidth]{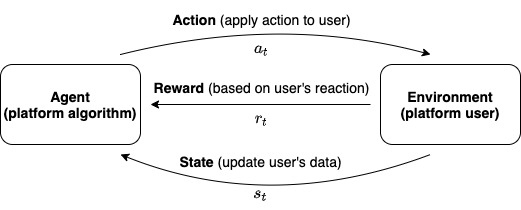}
    \caption{Schematic of reinforcement learning as applied to users of an internet platform.}
    \label{fig:RL}
\end{figure}

Different from supervised learning algorithms that learn from a pre-existing set of labeled data (\emph{offline learning}), RL takes an \emph{active learning} approach  where an \emph{agent} actively collects data by interacting with an unknown dynamic environment. In the case of RL on platforms, the unknown dynamic environment is the user. RL formalizes human-machine interaction histories as sequences of \{state, action, reward\} trajectories, generated while interacting in real-time with users as well as from previously collected user interactions \citep{sutton2018reinforcement}. The RL agent's goal is to intervene in its environment of human users to learn an optimal \emph{policy} maximizing the accumulation of designer-specified rewards (e.g., clicks). 
Figure \ref{fig:RL} illustrates the operation of RL in a platform context.

RL is considered a machine learning approach because the agent takes a series of decisions to maximize the cumulative reward for a predefined task without being explicitly programmed to achieve the task. In other  words, the goal is for the agent to learn the optimal behavior through repeated trial-and-error interactions with the environment, without human involvement \citep{matlab}. RL is more suitable than human-designed randomized experiments (e.g. A/B testing) when the space of possible interventions is huge, and there is no known functional form relating the outcome to the inputs.
The platform environment is exactly of the latter type, where the number and type of potential BMOD interventions is extremely high: there is a wide range of different content that can be displayed\footnote{``People who do a lot of research on products may see an ad that features positive product reviews, whereas those who have signed up for regular deliveries of other products in the past might see an ad offering a discount for those who “Subscribe \& Save.”" \url{www.nytimes.com/2019/01/20/technology/amazon-ads-advertising.html}} (e.g.~ads, news items, friend suggestions), a variety of ways to serve the content (e.g., format, timing, device), and a variety of positive and negative reinforcement types (e.g.~positive reinforcement with rewards, recognition or praise; or negative reinforcement with time pressure or social pressure). These can further be personalized 
by utilizing users' traits combined with their implicit feedback \citep{denreinforcement}. For example, \citet{kosinski2013private} showed how Facebook users' Likes can predict their psychological attributes, ranging from sexual orientation to intelligence, and suggested that including such attributes can improve personalized interventions.\footnote{``online insurance advertisements might emphasize security when facing emotionally unstable (neurotic) users but stress potential threats when dealing with emotionally stable ones." \citep{kosinski2013private}}. 
With such a large space of interventions (with potentially some being continuous), platforms would be unwilling or unable to learn offline models for the entire space.
And indeed, RL algorithms govern personalized interventions on many platforms. For example, recommender systems used by popular commercial platforms such as TikTok, Pandora, Instagram, and YouTube use interactive user data to select an optimal recommendation policy for a given user~\citep{zhou2020interactive, chen2019top}. Facebook uses RL to modify a user’s likelihood of clicking a push notification \citep{gauci2018horizon}. LinkedIn uses multi-armed bandits, a simplified type of RL, for automating ad placement decisions \citep{tang2013automatic}, and Yahoo! uses contextual multi-armed bandits for personalized news recommendations \citep{li2010contextual}.

In short, RL has three features that make it a useful approach for ``improving" prediction: It employs interventions, it learns online (as opposed to modeling only pre-existing data), and it is based on delayed reward (allowing it to sequentially improve).
The ``improve" prediction strategy we've described will occur if {during the period $(t,t+k)$ RL is deployed online to the showcase sample, with an} objective function set to minimizing prediction error where the predictions were already generated {at time $t$ (}prior to the BMOD{)}\footnote{{While the RL could potentially use offline data prior to time $t$ to learn the policy for users in general, such offline learning would likely have a different objective function—such as maximizing user engagement—and would require a large set of users beyond the limited showcase sample. Moreover, it would be unable to exploit the feedback of individual users’ idiosyncratic responses to online interaction with the RL agent.}}. In other words, at time $t$, predictions are generated from a model that learned offline prior to time $t$. The RL's objective function is then set based on these predictions, leading the RL agent to (sequentially) modify behavior towards those predictions in the period $(t,t+k)$. At time $t+k$, the resulting users' behaviors will be closer to their predictions, resulting in apparent ``improved" predictive performance. As mentioned earlier, setting the RL's objective function in this way can be done maliciously, myopically, or even erroneously. 

\subsection{A predict-then-modify scenario}\label{subsec-scenario}

The following (fictitious) scenario describes the different steps of a platform's engagement in the predict-then-modify strategy for the purpose of selling a prediction product. Consider a ride-sharing or social media platform that wants to sell predicted risk scores of drivers to an insurance company. The insurance company plans to buy such predictions on an ongoing basis. But first they want information about the expected prediction accuracy of the platforms’ predicted driver risk scores for comparing across platforms or for determining whether the accuracy would be sufficient for their purposes.
To that end, the platform's data science team operationalizes risk in a way that is measurable and acceptable to the insurance company. For example, they use “app usage rate while driving” ($y$) as the behavior of interest (assuming a high rate is indicative of higher risk). They also agree on the forecast horizon $k$ of interest, such as ``predicted risk in $k=7$ days". Importantly, we assume the platform's BMOD is capable of increasing/decreasing the app usage rate, for instance by controlling the level of engaging the driver on the app by varying the volume, type, or delivery mode of notifications.
The platform's proof-of-concept process would continue as follows:
\begin{description}
\item[Step 1: Get prediction model (\emph{Train}).] The data science team identifies a pre-trained prediction model with horizon $k$, $\hat{y}_{\tau+k}=\hat{f}(\bm{x_\tau})$, that was trained earlier using data from their platform (or can be quickly retrained on an appropriate subset of drivers' data relevant to the insurance company, such as on a restricted geographical area). 
\item[Step 2: Select showcase sample and generate predictions (\emph{Predict}).]  At time $t$, the platform selects a sample of $n$ drivers, the \emph{showcase sample}, for whom it has data $\bm{x_t}$. For each driver $i$, it predicts their app usage rate at time $t+k$ ($\hat{y}_{i,t+k}$). These predictions can immediately be shared with the insurance company.
\item[Step 3: Deploy BMOD (\emph{Modify}).] During the period $(t,t+k)$, the platform deploys BMOD sequentially to push the showcase drivers' behavior towards $\hat{y}_{t+k}$. At any point in time $t'\in (t,t+k)$, the platform (i.e.~RL agent) has information about $\hat{y}_{t+k}, \bm{x}_t{, y_t}$ and any further information that becomes available until $t'$. This step can be implemented via RL 
with an objective function of minimizing the deviation between the observed app usage rate and its corresponding (fixed) prediction $\hat{y}_{t+k}$. The RL agent utilizes the feedback afforded by the real-time stream of prediction error data given as the differences between $\hat{y}_{i,t+k}$ and $y_{i,t'}$, to continually nudge the use by varying the volume, type, or deliver mode of notifications. 
\item[Step 4: Evaluate predictive accuracy (\emph{Evaluate)}.] At time $t+k$, the behavior outcome of interest, app usage rates, is recorded for the showcase sample. These values are then compared to their predictions, and performance is summarized via metrics such as mean square error (MSE). This performance is shared with the insurance company. 
\end{description}
At this point, the insurance company evaluates the predictive performance and decides whether to purchase the prediction product. A purchase decision will result in the platform selling predicted scores for other drivers beyond the showcase sample, perhaps on an ongoing basis.

The effect of this predict-then-modify strategy on drivers is harmful: it would turn high-risk predictions into high-risk realities. We can envision other prediction products involving risk prediction that lead to further dangerous human and social outcomes. One example is a security or social services agency interested in purchasing predicted criminal behavior scores. The platform's proof-of-concept process could nudge high-scored platform users towards criminal behavior. Another example is a government interested in predicting citizens likely to become severely ill with Covid-19, for purposes of public health decision making. A platform showcasing their “Covid-19 susceptibility” scores product would push users towards or away from infection by encouraging or discouraging physical proximity with others and travelling to crowded location.
Note that in these three scenarios, the gap between the platform's goal (showcasing accurate predictions) and the customer's objective (avoiding high risk drivers, decreasing criminal behavior, and reducing Covid-19 harm) leads to pushing high-risk users' behaviors in a direction that is not only ethically dubious but also at odds with the customer's interests.

\subsection{Dataset shift that reduces prediction error} 
When prediction is not followed by behavior modification, differences between the training and deployment environments introduce uncertainty, typically by increasing bias and/or changing the noise distribution.
As described earlier, such dataset shift challenges can cause larger prediction errors at deployment, and are therefore a major challenge in adversarial attacks and gaming scenarios, which involve mischief during training or prediction \citep{huang2011adversarial}. By contrast, in behavior modification, training and deployment 
are made \emph{closer} by design.
While dataset shifts arising from uncontrollable and unforeseeable conditions increase uncertainty, behavior modification intends to shift actual outcome values $y$ \emph{closer} to $\hat{y}$ thereby reducing uncertainty. For example, when predicting that a user is likely to become depressed, displaying depressing news, friends' posts, and depression-related ads increases that user's chance of depression Facebook's emotional contagion experiment by \citet{kramer2014experimental} displays such capability.
When predicting the arrival time of a delivery, incentivizing faster or slower driving can increase the accuracy of the predicted arrival time. Displaying donation amounts by friends with amounts similar to the user's predicted amount can increase the chance the user donates the predicted amount \citep[e.g.,][]{nook2016prosocial}.

\subsection{``Improve" prediction vs.~adversarial attacks and gaming}
We note that the predict-then-modify strategy, which combines prediction and behavior modification, differs from {\it adversarial attacks}, where an attacker interferes in the process of training or prediction, manipulating the training data, prediction model, or predicted values~\citep{huang2011adversarial}. It also differs from gaming by strategic users, who manipulate their own behavior, the input data, to obtain a favorable prediction \citep[e.g.][]{hardt2016strategic,munro2020learning,frankel2019improving}. The first key difference is that adversarial attacks and gaming are aimed at modifying {\it predictions}, whereas the strategy we describe modifies the {\it actual behavior}. In the scenario we describe, rather than directly manipulating data, the platforms influence the behavioral processes that generate the data. The second key difference is that, in adversarial attacks and gaming, the intervention is performed by an attacker or a user, whereas in our case both prediction and behavior modification are performed by the platform.

\subsection{``Improve" prediction vs.~ causal classification}
{The predict-then-modify strategy differs from individual treatment assignment {that occurs in} causal classification, {that is,} identifying individuals whose outcome would be positively changed by a treatment, 
such as {in} uplift modeling \citep{fernandez2020comparison,olaya2020survey}. }
{A key difference is the sequence of actions: in causal classification prediction follows an intervention, whereas in the predict-then-modify scenario prediction precedes behavior modification. The business goals are vastly different: causal classification aims {at effective precision targeting,} 
whereas predict-then-modify uses causal manipulation to make the results of a prediction appear more accurate than it otherwise would have been.}

\section{Formalization and Analysis} 
\label{sec-formalization}
To study prediction error under the prediction-based behavior modification strategy, it is useful to  decompose the new form of expected prediction error (under behavior modification) into separate meaningful sources. This helps identify components such as bias, variance, and noise. However, we need a technical vocabulary that can encode both correlation-based prediction and causal-based actions.

The challenge is that standard notation and terminology used in statistics and machine learning for predictive modeling is insufficient for formalizing the problem of minimizing prediction error by intentionally manipulating the actual outcome values by way of behavior modification. The bottom panel of Figure \ref{fig:error-diagram} illustrates this new scenario. Specifically, predictive terminology conveys correlation-based relationships, but is not well-suited for denoting intentional manipulations. 
Figure \ref{fig:error-diagram}, which includes both causal arrows (orange) and a correlational connector (depicted as a squiggly black arrow, but with no causal interpretation\footnote{We chose to use a single-headed squiggly arrow rather than a bi-directional straight arrow for the correlation-based predictive relationship to convey the asymmetric input-output roles of $X$ and $y$. In standard causal diagrams, bi-directional arrows convey an unobservable variable affecting the two variables at the arrowheads, and there is no way to represent an asymmetric correlation-based predictive relationship.}) is incoherent in the world of causal diagrams, as well as in the world of prediction. {While the combination of prediction and causal intervention is present in the machine learning areas of causal classification (e.g. uplift modeling) and recommendations, until recently the literature in those areas either did not use causal language at all, or else it used causal language in an informal way \citep{zhang2021unified}. 
Only recently have papers started to appear that use formal causal notation along with predictive notation in a unified way, such as 
\citet{gutierrez2017causal,olaya2020survey,zhang2021unified,fernandez2020comparison,fernandez2022causal,verbeke2022or}. With the exception of \citet{olaya2020survey}, all these papers use the potential outcomes framework by \citet{rubin1974estimating}. Using a unified, formal notation for the prediction and causal components to represent scenarios that combine both actions is important for developing theory, deepening insight, and generalizing methodology to more complex scenarios\footnote{{We thank Wouter Verbeke for this point.}}.} 

{In line with this effort to provide a unified notation that combines prediction and intervention, we propose formalizing the predict-then-modify scenario by} integrating causal notation into existing predictive terminology in a parsimonious way. We do this by adopting the $do(.)$ operator by \citet{pearl2009causality}, where $do(B)$ denotes that variable $B$ is not simply observed but rather intentionally modified.\footnote{``The $do(x)$ operator is a mathematical device that helps us specify explicitly and formally what is held constant, and what is free to vary" \citep[][p. 358]{pearl2009causality}} This allows us to incorporate intentional behavioral modification into the predictive modeling context. We then use this notation to decompose the expected prediction error, to identify the different components affecting predictive power.
 
\subsection{Notation}
In the following we drop the time subscripts used in Section \ref{subsec-scenario} for a parsimonious exposition. We also use a continuous $y$.

For behavior prediction (steps 1-2) we use the ordinary notation $\hat{y}$ to denote the predicted user behavior, given the user's profile $\bm{x}$ (which can also include contextual information such as location and time-of-day). The predictions are generated by a predictive {model} aimed at capturing the predictive (correlation-based) relationship between the outcome and the predictors from a training dataset:
\begin{equation}
   y | \bm{x} =  f(\bm{x}) + \epsilon.
   \label{eq:y_unmod}
\end{equation}

Using the trained predictive {model} $\hat{f}$, trained offline using preexisting data ($\bm{x},\bm{y}$), a single-valued {$k$-step ahead} point prediction can be generated {at time $t$,} for each observation $i$ in the showcase sample: 
\begin{equation}
    \hat{y}_{i{,t+k}} | \bm{x}_{i{,t}} = \hat{f}(\bm{x}_{i{,t}}).
\end{equation}

Behavior modification (step 3) requires introducing interventional notation. We use $do(B)$ to denote the intentionally-modified platform behavior. Specifically, we use Pearl's ``intervention as variable" formulation  \citep[][Section 3.2.2]{pearl2009causality} that conceptualizes the intervention as an external force that alters the \emph{function} between $B$ and $X$.
The important advantage of this formulation (over the simpler conceptualization of forcing $B$ to take on a fixed value) is that ``it contains information about richer types of interventions and is applicable to any change in the function relationship $f_i$ and not merely to the replacement of $f_i$ by a constant" (p. 71).

\begin{figure}[]
    \begin{subfigure}{\textwidth}
        \centering
        \includegraphics[width=0.35\textwidth]{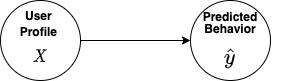}
         \caption{Generating a predicted value}
    \end{subfigure}
    \begin{subfigure}{\textwidth}
        \centering
        \vspace{0.3in}
        \includegraphics[width=0.5\textwidth]{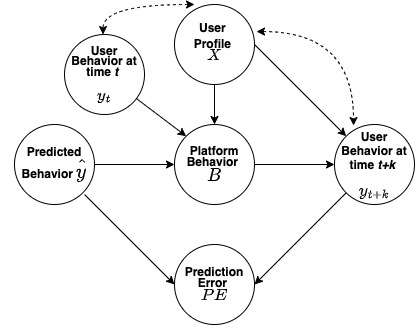}
         \caption{Modifying behavior towards the predicted value}
    \end{subfigure}
     \caption{Causal diagrams representing the separate steps of (a) generating a predicted value at time $t$  and (b) modifying behavior towards the predicted value during $(t,t+k)$. The doubled-headed dotted arrow{s} represent correlation between $X$ and $y$ that occurs by means of their being influenced by a common set of unobserved or latent variables.}
    \label{fig:dags}
\end{figure}

Figure \ref{fig:dags} illustrates the two separate steps of first generating a prediction at time $t$, and then modifying behavior toward the prediction during $(t,t+k)$, using causal diagrams. Nodes represent the different variables. The arrows mean that the node with incoming arrow(s) is a function of the nodes pointing to it. Note that in diagram (b) there is no arrow into $\hat{y}$ because at this stage (during $(t,t+k))$ it is assumed fixed.

Next, we denote the manipulated {behavior} as $\tilde{y}$. 
We note that it is incorrect to write $\tilde{y}\doteq do(y)$ because the user's outcome $y$ is not directly manipulated. Instead, the modified behavior is fully mediated: the platform tailors its behavior $B$ ($do(B)$ or personalized $do(B_i)$) to manipulate the user's mental state (emotion, feeling, mood, thought, etc.), which in turn leads to the modified behavior $\tilde{y}_i$. This modification is specifically aimed at pushing the outcome towards its prediction, and thus $do(B)$ inherently uses $\hat{y}$. Using $\sim$ on top of terms affected by $do(B)$, we therefore write\footnote{Expressions using the $do(.)$ operator are typically written within causal queries expressed in the form of conditional expectations or probabilities, such as $E[y_i|do(B_i), x_i]$. 
We take some license to show it outside of that usual casing because our focus here is on representing the manipulated outcome rather than estimating a causal effect.}
\begin{equation}
\label{eq:y_mod1}
\tilde{y}_{i{,t+k}} \doteq y_{i{,t+k}}|do(B_i),\bm{x}_{i{,t}},
\end{equation}
where $B_i = h(\hat{y}_{i{,t+k}}, \bm{x}_{i{,t}},{y_{i,t}})$.\footnote{{For simplicity, we assume a deterministic intervention $B_i$. However, our results Section \ref{subsec-EPE-tilde} are valid also for a stochastic modification of the form $B_i = h(\hat{y}_i, \bm{x}_i) + \upsilon_i$ where $\upsilon_i$ is an error term. Stochastic interventions {could reflect effects that are beyond the platform's control, such that the treatment a user receives may differ slightly from the platform's intended treatment. {For example, when the intervention is a displayed news item or ad, its choice might be affected by} the posting activity of a user's friends, unexpected events in the news, the stock market, the weather, or the time(s) of day that the user may choose to log into the app. All of these could affect the news or ads viewed by the user, such that the firm may have intended a slightly different level in quantity or quality of treatment than what the user actually received. {Moreover,} 
the firm could decide to intentionally build in some noise into {personalized} treatments.}}} 
Including $\bm{x}_{i{,t}}$ {and $y_{i,t}$} in $h(\cdot)$ represents personalized behavioral modification, based on the user's specific predictor information $\bm{x}_i$ (e.g.,~user $i$'s browsing history, demographics, location) {and their outcome at time $t$}. Note that $\bm{x}_{i,t}$ {and $y_{i,t}$} can impact the modified behavior both directly and indirectly via the platform's personalized BMOD. In fact, $\bm{X}_i$ is a collection of variables; some might be used to personalize $B_i$ and others (or even the same ones) moderate the effect of $B_i$ on $\tilde{y}_{i{,t+k}}$.
These causal relationships are graphically depicted in the bottom diagram in Figure \ref{fig:dags}.

For the modified outcome, we use $f_{do}$ to denote the underlying function, which can  be a completely different function from $f$:
\begin{equation}
\tilde{y}_{i,{t+k}} = 
f_{do}(do(B_i),\bm{x}_{i{,t}}) 
+ \tilde{\epsilon_i} = g(do(B_i),\bm{x}_{i{,t}}) 
+ \tilde{\epsilon_i},
  \label{eq:y_mod_func}
\end{equation}
where {we assume that} $\tilde{\epsilon}_i$ has mean $0$ and variance $\tilde{\sigma}^2.$

We note that the quantity $E[\tilde{y}_i|\bm{x}_i] - E[y_i|\bm{x}_i] = E[\tilde{y}_i - y_i |\bm{x}_i]$, is called the (population) \emph{Conditional Average Treatment Effect} (CATE)  \citep{athey2016recursive,imbens2015causal} or \emph{Individual Treatment Effect} (ITE) \citep{shalit2017estimating} and is of key interest in treatment effect estimation and testing.\footnote{Note that $y_i|\bm{x}_i$ assumes no manipulation. \citet[][p. 70-72]{pearl2009causality} offers an alternative formulation to encode manipulation vs.~no manipulation, by adding a binary intervention indicator $I_B$ that obtains values in $\{do(B_i),idle\}$. In our case $I_B=idle$ for $y_i|\bm{x}_i$.}

The prediction errors at time $t+k$ are obtained by comparing the predicted values $\hat{y}{_{t+k}}$ to the manipulated outcomes $\tilde{y}{_{t+k}}$.
Table \ref{tab:notation} provides the short notation, full notation and description for each of the above terms.\footnote{It is possible to use Rubin's potential outcomes notation intended for estimating treatment effects \citep[e.g.][p. 33]{imbens2015causal}. This requires defining $B=\{0,1,2,\ldots\}$ as the intervention assignment, and denoting by $y_i(B)$ the outcome, where $y_i(0)$ is the un-manipulated outcome. The quantity $y_i|do(B),\bm{x}$ is written as $y_i(B)|B,\bm{x}$. We prefer the $do(.)$ operator since it conveys the causal nature of the manipulation $B$ and clearly differentiates it from the correlation-based prediction components $\bm{x}$. Another possibility is using the counterfactual notation in \citet[][chapter 4]{pearl2016causal} recently used in algorithmic fairness research, e.g. \citet{zhang2018fairness,kusner2017counterfactual}. While the counterfactual notation is more general than the $do(.)$ operator, since we focus on interventions that do not require counterfactual notions, the $do(.)$ operator provides a more parsimonious and elegant choice.} {For clarity, we sometimes drop subscripts $i$ or $t$ and $t+k$ in the following {exposition}.} Together with Equations (\ref{eq:y_mod1})-(\ref{eq:y_mod_func}), we now have a sufficient vocabulary for examining the prediction error under behavior modification. 
\begin{table}
    \caption{Short and full notation}
    \label{tab:notation}
    \centering
    \begin{tabular}{|c|l|l|}
\hline
    Short  & Full notation/definition & Description \\
    notation & & \\ \hline 
    $y_i$   & $y_i|\bm{x}_i = f(\bm{x}_i) + \epsilon_i$ 
    & Outcome under no manipulation \\    
    $f$     & $f(\bm{x}) = E[y|\bm{x}] $ & True function under no manipulation \\
    $\sigma^2$ & $Var(\epsilon)=E[\epsilon^2]$ & Noise variance under no manipulation \\
    $\hat{y}_i$   &$\hat{y}_i| \bm{x}_i  = \hat{f}(\bm{x}_i)$ & Predicted outcome under no manipulation \\
    {$\hat{y}_{i,t+k}$}   & { $\hat{y}_{i,t+k}| \bm{x}_{i,t}  = \hat{f}(\bm{x}_{i,t})$} & {Predicted $k$-step ahead outcome under no manipulation} \\
    $B_i$ & $h(\hat{y}_{i{,t+k}},\bm{x}_{i{,t}},{y_{i,t}})$ 
    & Personalized intervention \\
  $f_{do}$ & $g(do(B),\bm{x})= E[y| do(B),\bm{x}] $ 
  & True function under $do(B)$ \\
  $\tilde{y}_{i{,t+k}}$ & $y_{i{,t+k}}|do(B_i),\bm{x}_{i{,t}}
  = g(do(B_i),\bm{x}_{i{,t}})
  + \tilde{\epsilon_i}$
  & Manipulated outcome {at time $t+k$}\\
  $\tilde{\sigma}^2$ & $Var(\tilde{\epsilon})=Var(\tilde{y})=E[\tilde{\epsilon}^2]$ & Noise variance under $do(B)$ \\
\hline
    \end{tabular}
\end{table}

\subsection{Expected prediction error of modified outcomes (\texorpdfstring{$\widetilde{EPE}$}{Lg})}
\label{subsec-EPE-tilde}
When outcome values are intentionally pushed \emph{towards} their predictive values, it is intuitive that the resulting expected prediction error will be lower than the no-manipulation outcome values.\footnote{In the prediction minimization process \emph{all} subjects are initially not $B$-manipulated and later a sample is $B$-manipulated using personalized modifications.} 
We can now formalize the following questions: Given $\hat{f}$, a specific predictive {model} trained on data with no behavior modification ($X,\bm{y})$, when will the expected prediction error for a manipulated user with predictors $\bm{x}$ and manipulation $do(B_i)=b$ be lower than if the user was not manipulated? That is,
for {loss function $L$},  when will we get
\begin{equation}\label{eq:Lp}
{E[L(\tilde{y}, \hat{f}(b,\bm{x}))] < E[L(y , \hat{f}(\bm{x}))]} ?
\end{equation}
When might the manipulation lead to worse predictive power? 

To answer these questions, we proceed to break down the EPE into several non-overlapping components. 
Using the standard $L_2$ loss function, we can obtain the EPE under behavior modification as follows (the full derivation is given in \ref{append-decomposition}):
\begin{equation}
\begin{split}
\widetilde{EPE}(\bm{x}) &= E\left[\left(y|{do(B), \bm{x}} 
- \hat{f}(\bm{x})\right)^2 \right] \\
& = \tilde{\sigma}^2 + \left[CATE(\bm{x}) + Bias(\hat{f}(\bm{x}))\right]^2 + Var(\hat{f}(\bm{x})).
\end{split}
\label{eq:epe2}
\end{equation}
Each of the terms in Equation~(\ref{eq:epe2}) has an interesting meaning and different implications on the effect of behavior modification on EPE. The additive nature of this formulation provides insights on the roles of data size, predictive {model} properties, and behavior modification qualities. By comparing $\widetilde{EPE}(\bm{x})$ to $EPE(\bm{x})$ (the manipulated and non-manipulated scenarios), we can see the following:
\begin{itemize}
\item \emph{Data size:} Whether manipulating or not, data size affects $\widetilde{EPE}$ via the variance of {$\hat{f}$},\footnote{The machine learning \emph{bias} is asymptotic in sample size: an algorithm is biased ``if no matter how much training data we give it, the learning curve will never reach perfect accuracy" \citep{provost2013data}.} indicating that larger training samples can improve not only predictions, but also the average manipulated prediction error. Pushing the outcome towards a more stable prediction leads to smaller errors. 
\item \emph{Magnitude of behavior modification effect:} The second term shows the role of the average behavior modification magnitude (CATE) in countering the bias of $\hat{f}$. 
This term is minimized when $CATE=-Bias(\hat{f})$, that is, when, on average, $do(B)$ pushes the user's behavior in a direction and magnitude that exactly counters the bias of {$\hat{f}$}.  
Thus, an effective behavior modification can improve predictive power by combating the bias in {$\hat{f}$}, as long as $0 < CATE < -2Bias$ or  $-2Bias < CATE < 0$. 
\item \emph{Noise (homogeneity of prediction errors):} Compared to $\sigma^2$ in the no-manipulation $EPE$, the first term in $\widetilde{EPE}$ is $\tilde{\sigma}^2$, the noise variance \emph{under behavior modification}. This means behavior modification can also affect the \emph{variability} of prediction errors across different users.
\end{itemize}

\subsection{Modification Strategies}
We now examine the trade-offs and implications of the four $\widetilde{EPE}$ sources ($\tilde{\sigma}$, $CATE$, $Bias (\hat{f})$, $Var(\hat{f})$) on the expected prediction error. The resulting insights help determine ideal modification under different scenarios of algorithm bias and variance.

\begin{figure}[]
    \centering
    \includegraphics[width=\textwidth]{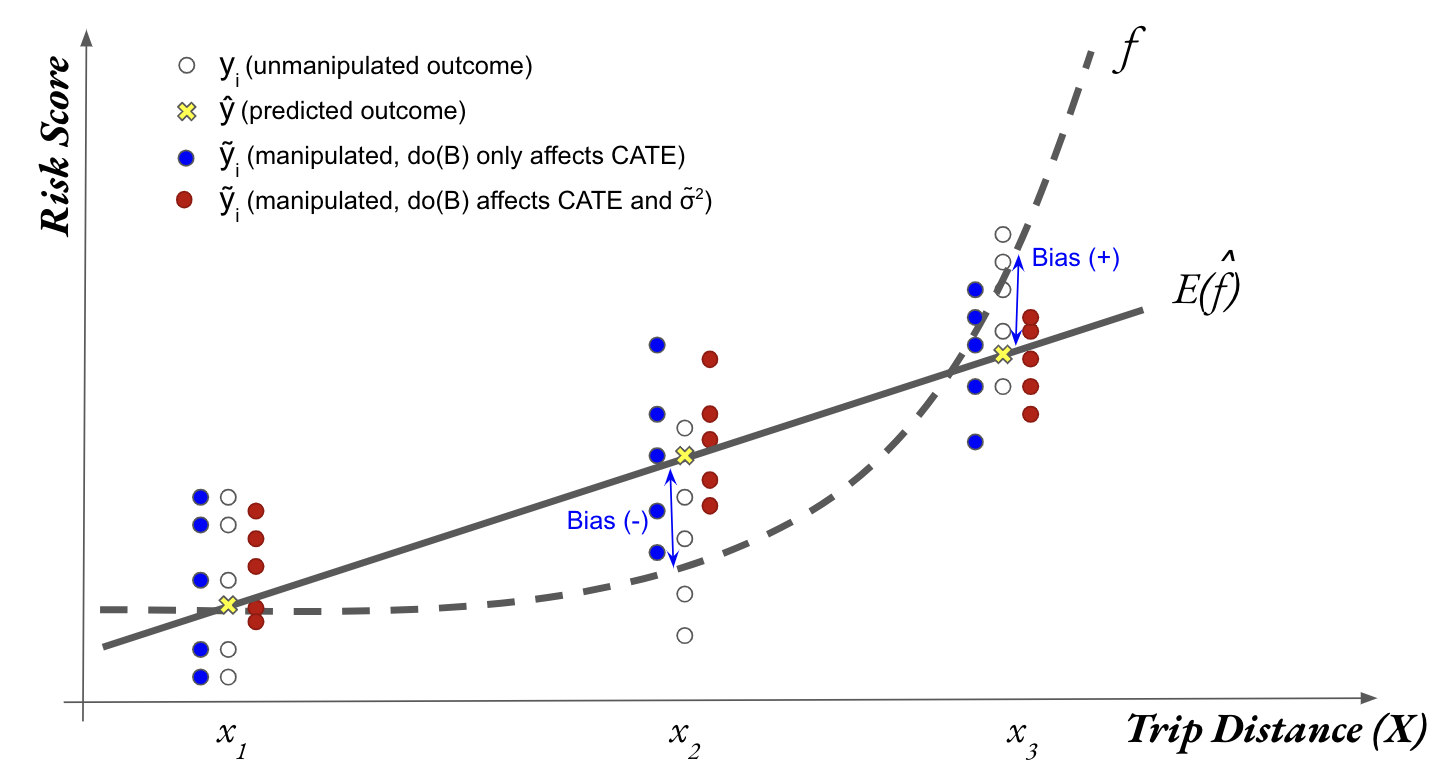}
    \caption{Hypothetical prediction of risky driving behaviors given distance, by ride-sharing platform. Illustrates the effect of behavior modification on shifting the average outcome by $CATE=-Bias$ (blue circles) or on both shifting the average outcome and shrinking the variance $\tilde{\sigma}^2$ (red circles). Yellow `X' symbols represent predicted values $\hat{f}(x_i)$. (The schematic assumes a very large training sample, and thus $\hat{f}\approx E[\hat{f}]$.)}
    \label{fig:bias-var-cate}
\end{figure}

To this end, we return to the ride-sharing example from Section \ref{subsec-scenario}, where a ride-sharing or social media platform is showcasing its risk prediction capability to an insurance company. The platform can modify drivers' behaviors by manipulating the driver's engagement with their app while driving. Figure \ref{fig:bias-var-cate} shows a scatter plot of driver risk score (y-axis) vs. daily distance (x-axis) for a fictitious showcase sample of drivers (hollow circles). Suppose predictions are drivers' 
predicted rates of app usage while driving (\emph{risk scores})
and the goal is to minimize the (squared) differences between the predicted and actual values.
Suppose that risk is a quadratic function of driving distance, but that the predictive model estimates a linear relationship. The linear model's predicted risk scores are shown as yellow `x' in Figure \ref{fig:bias-var-cate}. We consider three specific distances: $x_1, x_2,$ and $x_3$. While $\hat{f}$ is biased, for $x_1$ there is no bias, for $x_2$ the bias is negative, and for $x_3$ bias is positive (for simplicity, the schematic assumes a very large training sample and thus $\hat{f}\approx E[\hat{f}]$). Next, the blue and red circles represent two types of BMOD: the first BMOD affects only $CATE$ while the second affects both $CATE$ and $\tilde{\sigma}$.

\subsubsection*{Scenario 1: Low-bias $\hat{f}$ trained on a very large sample.}
This scenario would be akin to deep learning algorithms applied to massive training data. The very large sample means $Var(\hat{f})\approx 0$ and $Bias(\hat{f})$ is very small. The strategy of setting $CATE= -Bias(\hat{f})$ is optimal if the behavior modification also {does not increase} error heterogeneity so that $\tilde{\sigma}^2 \leq \sigma^2$. Because the bias is low, the optimal behavior modification should have a small effect. In Figure \ref{fig:bias-var-cate},  $\hat{f}(x_1)$ has no bias, and therefore applying behavior modification to drivers with distance $x_1$ will introduce bias, and is only useful if it can sufficiently shrink the variability of the resulting risky behaviors
(red circles).

\subsubsection*{Scenario 2: High-bias $\hat{f}$ trained on a very large sample.}
{Algorithms that lead to $\hat{f}$ with} high bias 
include naive Bayes, linear regression, shallow trees, and k-NN with large $k$. As in scenario 1, here too $Var(\hat{f})\approx 0$. The strategy of setting $CATE= -Bias(\hat{f})$ is depicted in Figure \ref{fig:bias-var-cate} as increasing average risky behaviors for $x_2$ by $|Bias(\hat{f}(x_2))|$ and decreasing it for $x_3$ by $|Bias(\hat{f}(x_3))|$. This strategy is optimal if the behavior modification also decreases (or at least does not increase) error heterogeneity so that $\tilde{\sigma}^2 \leq \sigma^2$. While a small modification effect in the right direction can help counter bias, the ideal modification effect must be as large as the bias. 
Note that EPE is computed for a specific $\bm{x}$, and therefore generalizing the above rule to any $\bm{x}$ requires either assuming homoskedastic errors $\tilde{\epsilon}$, or that the inequality holds for all $\bm{x}$ ($\forall \bm{x} \: \: \tilde{\sigma}^2_{\bm{x}} \leq \sigma^2_{\bm{x}}$).

\subsubsection*{Scenario 3: High-variance $\hat{f}$.}
If the predictive model $f$ is {trained} on a relatively small sample, {and the algorithm used leads to high variance in $\hat{f}$,} then potential minimization of $\tilde{\sigma}^2$ and/or $\left[CATE+Bias(\hat{f})\right]^2$ by way of behavior modification might be negligible relative to $Var(\hat{f})$. Because behavior modification is based on {nudging} behavior towards $\hat{f}(\bm{x}_i)$, a highly volatile $\hat{f}$ might result in erratic $do(B_i)$ modifications in terms of magnitude or even direction. {A sequential BMOD, such as via an RL agent, would take longer to learn.}

\section{Discussion}
\label{sec-discuss}

We described a new strategy that {platforms can use} for reducing prediction error which is completely different from approaches taken by the fields of statistics and machine learning. {Our focus is on interventions implemented {after exhausting the possibilities of prediction using offline data} 
with the goal of making {such} predictions 
seem more accurate before they are used by a customer}. This strategy combines prediction with behavior modification, and therefore formalizing it into technical language requires supplementing predictive notation with causal terminology. Using the $do(.)$ operator, we are able to describe the entire system that includes the training dataset, the predictive {model}, and the behavior modification. 

While our $\widetilde{EPE}$ formula also applies to behavior modification for commercial benefit (e.g.~advertising), we have focused on the more general case which can lead to dangerous and perhaps unintentional outcomes that not only harm users but also platform customers. Such an outcome can be achieved intentionally or myopically by data science teams 
when the platform uses automated personalization algorithms, such as reinforcement learning with an objective function  set to minimizing some function of $\hat{y}-\tilde{y}$, the difference between predicted and manipulated outcomes. In such cases the platform and customer goals can be misaligned, as in risk prediction applications where the customers aims to reduce risk, while the platform pushes risky users towards risky actions.

\subsection{Technical and business implications}
The contrast between the bias-variance decomposition of the manipulated and non-manipulated scenarios highlighted two key sources of the manipulated prediction error: the CATE-bias relationship, and its tradeoff with the manipulated noise variance. We now use these insights to return to the questions we posed earlier.
 
 \begin{enumerate}
     \item \emph{Can behavior modification mask poor predictive performance?}
Behavioral big data is noisy, sparse, and high-dimensional \citep{de2020benchmarking}. Behavior modification can improve $\widetilde{EPE}$ by countering the bias {in $\hat{f}$} as well as by reducing the noise variance. This means that poor performance of a predictive {model}, due to {high} bias and{/or} variance, and/or due to the data noisiness, can be masked by $do(B)$. Therefore, platform customers wanting to achieve the (manipulated) prediction accuracy level demonstrated by the platform, must acquire both the predictions \emph{and} the ability to apply BMOD similar to the one performed by the platform. Purchasing the predictions alone might uncover a much weaker predictive performance when deployed to non-manipulated users (or by applying a less effective BMOD). 

\item \emph{Can one infer the counterfactual $EPE$ from the manipulated $\widetilde{EPE}$?}
The difference between the two quantities of no-manipulation $EPE$ and behavior-modified $\widetilde{EPE}$ involves $CATE$, $bias(\hat{f})$, $\sigma$, and $\tilde{\sigma}$.\footnote{$\widetilde{EPE}-EPE = \tilde{\sigma}^2-\sigma^2 + CATE^2 + 2\times CATE\times Bias(\hat{f})$} Some of these quantities can be estimated by the platform (e.g., CATE), while others are more difficult, if impossible, to estimate. Hence, it is unlikely the no-manipulation predictive power can be ascertained from the manipulated $\widetilde{EPE}$. This means platform customers who want to evaluate the no-manipulation predictive power will need to acquire information about the estimated $EPE$ at the time of {model} testing. 

 \item \emph{Can customers detect the manipulation via A/B tests?} This would requires the customer to know which of its users we in the platform's showcase sample, and A/B tests  
 are unlikely to detect the error minimization strategy, because of the random allocation of users in an A/B test. This randomization spreads {BMOD-affected} users across the A and B conditions, and therefore the difference between the group averages will cancel out the {BMOD} effect. The A/B test statistic and its statistical significance are therefore not impacted by BMOD.

 \item \emph{What are the roles of personalized prediction and personalized behavior modification in error minimization?}
 Personalized prediction plays an important role: the more accurate the prediction, the less BMOD is needed to reach good predictive accuracy. Specifically, low-bias algorithms trained on very large samples (to shrink $Var(\hat{f})$) are advantageous in terms of requiring a smaller BMOD effect to minimize EPE. Hence, platform efforts and investments in improving personalized predictions is warranted. At the same time, in the ``improve" prediction strategy, personalizing the BMOD serves a complementary role: An ideal {BMOD}  reduces not only the average magnitude of the errors but also their variability, so that errors are more consistent across users. This highlights the role of \emph{personalized BMOD} in which
companies or platforms may invest: utilizing a user's personal $\bm{x}_i$ data  to select the best modification {$do(B_i)=h(\hat{y}_i,\bm{x}_i)$}, among the very large space of potential $B$ interventions. 
    Personalized BMOD has the potential to minimize $\widetilde{EPE}$ more equally both within a certain user profile $\bm{x}$ and across different user profiles, by lowering conditional bias via manipulating CATE, and by shrinking the (manipulated) outcomes' variance. 
 \end{enumerate}

\subsection{Humanistic and societal implications}
Behavior modification, now pervasively applied by platforms to their ``data subjects", is geared towards optimizing the platform's commercial interest, often at the cost of users' well-being and agency. AI expert \cite{russell2019human} highlights the advantage of making users' behavior more predictable:
``A more predictable user can be fed items that they are likely to click on, thereby generating more revenue. People with more extreme political views tend to be more predictable in which items they will click on."

\emph{Persuasive technology}, a design philosophy now implemented on platforms from e-commerce sites and social networks to smartphones and fitness wristbands, aims at generating \emph{behavioral change} and \emph{habit formation}, most often without users' knowledge or consent \citep{rushkoff2019team}.
This application of behavior modification to platform \emph{users} is more extreme than an organization applying BMOD to its \emph{employees} for increasing the organization's productivity. 
And, clearly, such use diverges from the original intention of behavior modification procedures ``to change socially significant behaviors, with the goal of improving some aspect of a person’s life" \citep[][p. 5]{miltenberger2015behavior}.

Given the often conflicting goals of data subjects and the platforms that collect and use their data as well as manipulate their behavior, it is important to introduce causal {notation} into the predictive environment, so that our statistics, machine learning, and computational social science communities  can study their technical properties and implications. By introducing and integrating causal notation into the predictive terminology, we can start studying how behavior modification can appear to create ``better" predictions. This allows examining the effects of different behavior modification types, magnitudes, variation, and directions on anticipated outcomes.

\section{Conclusion and Future Directions}
\label{sec-conclude}
Behavior modification can make users’ behavior not only more predictable but also more homogeneous. However, the apparent ``predictability" at the time of predictive evaluation ($t+k$) can be deceptive to the platform customers, to the extent that {predictive performance} based on manipulated behavior does not generalize outside of the platform environment. The apparent predictability at time $t+k$ also does not generalize within-platform, if the exact same BMOD is no longer used. This is because in a prediction product sales scenario, the pre-sale BMOD is a temporary action applied by the platform to a limited showcase sample and tuned to minimize a function of $\hat{y}-\tilde{y}$ for this showcase sample. Post-sale, it is reasonable that the exact same BMOD will no longer be used. Even if the platform's BMOD has long-term effects beyond time $t+k$, it would affect only the showcase sample but not new users scored for the prediction product. The latter are no longer subject to the predict-then-modify strategy. Hence generalizability is still compromised. Finally, the scenario where the platform continues to apply the same BMOD to users beyond the showcase sample, affecting users in the short or long term, is in itself alarming, as it raises concerns of whether the business customer is aware it is receiving an artificially altered prediction product and is complicit in this, or if it is being deceived.

Importantly, outcomes pushed towards their predictions may be at odds with the platform customers' interests, and harmful to the manipulated users. While platforms have the incentive and capabilities to minimize prediction errors, the predict-then-modify strategy is likely to occur, given the growing use of automated personalization techniques such as reinforcement learning, that interact with users, apply sequential interventions, and combine prediction and behavior modification. It is therefore critical to have a useful technical vocabulary that integrates intentional behavior modification into the correlation-based predictive framework to enable studying such contemporary strategies. We demonstrated how this notation can be useful to explore such issues as how behavior modification can mask poor predictive {power}, whether one {can} infer the counterfactual of non-manipulated predictive power from the manipulated predictive power, whether customers running routine A/B tests on the platform can detect behavioral modification schemes, and the possible role  of personalized behavioral modification schemes.

Future research can adopt the technical vocabulary and notation we proposed in this paper to further examine the potential scope and effects of behavioral modification by digital platforms. 
One such direction is the consideration of dichotomous outcomes. Our $\widetilde{EPE}$ derivation is for a continuous outcome. Deriving $\widetilde{EPE}$ for a binary outcome would be useful in applications where the behavior of interest is dichotomous, such as voting/not voting, responding/not-responding, or passing/failing some criterion. In such cases, the  predict-then-modify scenario would push the user's behavior towards the outcome with the higher predicted probability. There are different ways to formulate the EPE for binary predictions, as well as other ways to measure predicive accuracy. Results for the ordinary un-manipulated scenario show that
the effects of bias and variance on EPE are multiplicative rather than additive, and the literature reports conflicting results on their roles \citep[e.g.][]{friedman1997bias,domingos2000unified}. Formulating the effect of BMOD on the resultant prediction error in the binary outcome case can further highlight the distinction between estimating causal effects and selecting appropriate interventions \citep{fernandez2022ijds}. We leave derivations of such formulas for future research.

The scenario we've described is for a prediction product with a single prediction horizon $k$. An interesting future direction is considering a platform wanting to showcase predictions for multiple horizons simultaneously (e.g.~ “risk in the next day” and “risk in the next week”). In this case, an additional layer of complexity is added because aiming to apply BMOD for one horizon might affect the other horizon.

Another direction for expanding this work is using the notation, along with causal diagrams, to formalize and study \emph{non-platform} predict-then-modify scenarios, where the organization only generates predictions, but the prediction itself leads to behavior modification: in financial markets a stock recommendation affects stock price; in healthcare, a patient's predicted health outcome affects the behavior of those knowledgeable of the prediction (the patient, doctor, etc.) in a way that can eventually modify that outcome.

\bibliographystyle{apalike} 
\bibliography{predictionerror.bib}

\section*{Acknowledgements}
 We thank Foster Provost, Rob Hyndman, Pierre Pinson, Soumya Ray, Sam Ransbothan, Carlos Cinelli, Wouter Verbeke, Travis Greene, Boaz Shmueli, Patricia Kuo, Aurélie Lemmens, Rodrigo Belo, David Budescu, Raquelle Azran, Bill Meeker, and participants of 2020 SCECR, 2020 ENBIS conference, 2021 IIF Symposium, and seminars at the University of Iowa, Purdue, Erasmus University, Nova SBE, and University of Antwerp for invaluable feedback. We thank Noa Shmueli for Figure 1 artwork. G. Shmueli's research is partially supported by Ministry of Science and Technology, Taiwan [Grant 108-2410-H-007-091-MY3].

\newpage

\textbf{\large Appendix}
\appendix


\section{Derivation of the \texorpdfstring{$\widetilde{EPE}$}{Lg} Bias-Variance Decomposition (Equation (\ref{eq:epe2}))}\label{append-decomposition}

The derivation for Equation (\ref{eq:epe2}) (bias-variance decomposition under BMOD) is as follows.
For convenience, we use $f$ and $\hat{f}$ to denote the no-manipulation true function and its estimated model. For the manipulated scenario, we use $f_{do}$ to denote the true function under BMOD.

For a new observation with inputs $\bm{x}$ and manipulated outcome $y|do(B),\bm{x}$, we can decompose the expected prediction error as follows (for convenience, we drop subscripts $i${, $t$ and $t+k$}):
\begin{equation}
\begin{split}
\widetilde{EPE}(\bm{x}) &= E \left[ \left(y|do(B),\bm{x} - \hat{f}(\bm{x})\right)^2 \right]\\ 
&= E\left[ (\tilde{y}-\hat{f})^2 \right]\\
&= E\left[ (\tilde{y}- f_{do} + f_{do} - \hat{f})^2 \right]\\
& = E\left[(\tilde{y}-f_{do})^2\right] + E\left[(f_{do}-\hat{f})^2\right] + 2E\left[( \tilde{y}-f_{do})(f_{do}-\hat{f})\right]. 
\end{split}
\label{eq:epe1}
\end{equation}
These three terms can be further simplified.
The first term can be simplified by noting that $\tilde{y}=f_{do}+\tilde{\epsilon}$:
\begin{equation}
 E\left[(\tilde{y}-f_{do})^2 \right]  = E[\tilde{\epsilon}^2] = \tilde{\sigma}^2.
\end{equation}
The second term can be written as:
\begin{equation}\label{eq:10}
\begin{split}
   E\left[(f_{do}-\hat{f})^2 \right]&
= E\left[ \left( f_{do}-E(\hat{f}) + E(\hat{f}) -\hat{f} \right)^2 \right] \\ 
   & = \left( f_{do}-E[\hat{f}] \right)^2 + E\left[\left( \hat{f} - E[\hat{f}] \right)^2 \right]\\
    & = \left( f_{do}-E[\hat{f}] \right)^2 +Var(\hat{f})\\
\end{split}
\end{equation}
because the cross product is zero:
\begin{equation}
2 E\left[f_{do}-E[\hat{f})\right] \left( E[\hat{f}]-\hat{f} \right)  = 
2 \left(f_{do}-E[\hat{f}]\right) \left( E[\hat{f}]-E[\hat{f}] \right) =0.
\end{equation}
We can further write Equation \ref{eq:10} as a function of the bias and variance of $\hat{f}$:
\begin{equation}
    \begin{split}
 \left[ f_{do} - E[\hat{f}] \right]^2 + Var(\hat{f}) 
 &  = E\left[ f_{do} - f + f - E[\hat{f}] \right]^2 + Var(\hat{f}) \\
 &  = \left[CATE + Bias(\hat{f})\right]^2 + Var(\hat{f}). \\
\end{split}
\end{equation}

Finally, using the independence of the new observation's prediction error $\tilde{\epsilon}$ from the prediction $\hat{f}$ based on the training data ($E[\tilde{\epsilon}\hat{f}]=0$),  the third term can be shown to be zero:
\begin{equation}
2E\left[(\tilde{y}-f_{do})(f_{do}-\hat{f})\right]  = 2E[\tilde{\epsilon}](f_{do}-\hat{f}) = 2f_{do} E(\tilde{\epsilon}) - 2E[\tilde{\epsilon}\hat{f}] = 0.
\end{equation}

Therefore, we can write $\widetilde{EPE}$ from Equation (\ref{eq:epe1}) as:
\begin{equation}
    \begin{split}
    \widetilde{EPE}(\bm{x}) &= E \left[\left(y|\bm{x},do(B) - \hat{f}(\bm{x})\right)^2 \right]\\
    & =\tilde{\sigma}^2 + \left[CATE + Bias(\hat{f})\right]^2 + Var(\hat{f}).
    \end{split}
\end{equation}

\end{document}